\documentclass[a4paper]{article}

\usepackage{amsfonts}

\usepackage{slashed}

\def\be{\begin{equation}}
\def\ee{\end{equation}}
\def\bea{\begin{eqnarray}}
\def\eea{\end{eqnarray}}
\def\({\left(}
\def\){\right)}
\def\<{\left<}
\def\>{\right>}

\def\>{\rangle}
\def\<{\langle}
\def\|{\mid}

\def\tr{{\mbox{tr}}}
\def\be{\begin{equation}}
\def\ee{\end{equation}}
\def\bea{\begin{eqnarray*}}
\def\eea{\end{eqnarray*}}
\def\ben{\begin{eqnarray}}
\def\een{\end{eqnarray}}
\def\({\left(}
\def\){\right)}
\def\<{\left<}
\def\>{\right>}

\def\[{\left[}
\def\]{\right]}

\def\+{\bar}
\def\mb{\mathbb}
\def\tr{{\mbox{tr}}}

\def\t{\tilde}

\def\t{\widetilde}

\def\N{{\cal{N}}}

\def\ee{\breve{e}}

\def\+{\breve{+}}
\def\-{\breve{-}}

\def\M{\underline{M}}
\def\N{\underline{N}}

\begin{document}
\setlength{\unitlength}{1mm}

\pagestyle{empty}
\vskip-10pt
\vskip-10pt
\hfill 
\begin{center}
\vskip 3truecm
{\Large \bf
A preliminary test of Abelian D4-M5 duality}\\
\vskip 2truecm
{\large \bf
Andreas Gustavsson\footnote{a.r.gustavsson@swipnet.se}}\\
\vskip 1truecm
{\it  Physics Department, University of Seoul, 13 Siripdae, Seoul 130-743 Korea}
\end{center}
\vskip 2truecm
{\abstract{We compute the partition function of five-dimension Abelian sYM including a graviphoton term on a five-torus. The result agrees with Abelian M5 brane partition function with zero characteristics on an associated six-torus.}}

\vfill 
\vskip4pt
\eject
\pagestyle{plain}

The partition function of a single M5 brane on topologically non-trivial six-manifold $M$, can be computed using holomorphic factorization \cite{Witten:1996hc}. The holomorphic factorization was carried out more explicitly, and yet applicable to an arbitrary six-manifold, in \cite{Henningson:1999dm}. Here the classical contribution to the partition function was obtained in terms of a Jacobi theta function $\Theta\[\begin{array}{c}
a\\
b
\end{array}\](Z)$ where $a_i$ and $b^i$ are vectors with entries being either $0$ or $\frac{1}{2}$ and $i = 1,...,b_3$ where $b_3 = \dim H^3(M)$ is the third Betti number. A unique partition function was thus not obtained. An explicit computation of the partition function on a flat six-torus has been carried out using the Hamiltonian formalism and without using the holomorphic factorization in \cite{Dolan:1998qk}. It was found that the partition function with zero characteristics, $a_i = b^i = 0$ is picked. The same result was later also obtained using the holomorphic factorization method in \cite{Gustavsson:2000kr} where again it was found that the theta-function  with zero characteristic is favoured due to modular invariance.

It has been clear for a long time that the M5 brane can be dimensionally reduced to a D4 brane by compactifying the M5 brane on a circle. But recently it has been conjectured that these two theories might even be dual \cite{Lambert:2010iw}, \cite{Douglas:2010iu}. As very preliminary test of this proposal we will compute the partition function of five-dimensional D4 theory on a five torus, and show that it agrees with the partition function $\Theta(Z):=\Theta\[\begin{array}{c}
0\\
0
\end{array}\](Z)$ of the M5 brane on a corresponding six-torus.

Let us briefly recall the computation of the partition function of an Abelian selfdual three-form on an euclidean six-manifold $M$ \cite{Henningson:1999dm}. Let us introduce the notation
\bea
(\omega,\eta) &=& \int \omega \wedge * \eta
\eea
for two forms $\omega$ and $\eta$ of the same rank. As the euclidean Action we take \cite{Henningson:2004dh}
\bea
S &=& \frac{1}{2\pi} (H,H)
\eea
and we assume that $H$ is a nonchiral closed three-form field. By Hodge decomposition this field can be expressed as
\bea
H &=& H^0 + dB + d^{\dag} C
\eea
where $H^0$ is harmonic, $B$ is a globally defined two-form. However, since 
\bea
(H,d^{\dag}C) &=& (dH,C)\cr
(H,d^{\dag}C) &=& (d^{\dag}C,d^{\dag}C) > 0
\eea
and we assume that $dH = 0$, we find that the most general Hodge decomposition is reduced to two terms,
\bea
H &=& H^0 + dB
\eea
We will refer to the harmonic term as the classical part, and the globally exact term as a quantum fluctuation. Indeed $(H,H) \geq (H^0,H^0)$ as a consequence of $(H^0,dB) = 0$. Thus the classical part is a local minimum of the Action and vice versa. 

Dirac charge quantization means that if $a$ is any three-cycle in $M$, then 
\bea
\int_a H &\in & 2\pi {\mb{Z}}
\eea
We wish to expand the harmonic part in the basis $\{E_{Ai},E^i_B\}$ of harmonic three-forms 
\bea
H^0 &=& n^{Ai} E_{Ai} + n^B_i E_B^i
\eea
with integer coefficients $n^{Ai}$ and $n^B_i$. This is possible only if we restrict our basis to also have periods which are $2\pi$,
\bea
\(\begin{array}{cc}
\int_{a^i} E_{Ak} & \int_{a^i} E_B^l\\
\int_{b_j} E_{Ak} & \int_{b_j} E_B^l
\end{array}\) &=& 2\pi\(\begin{array}{cc}
\delta^i_k & 0 \\
0 & \delta_j^l
\end{array}\)
\eea
where $a^i$ and $b_j$ are three-cycles in $M$. Then we have the following intersection form of these three-cycles 
\bea
\(\begin{array}{cc}
\int E_{Ai} \wedge E_{Aj} & \int E_{Ai} \wedge E_B^j\\
\int E_B^i \wedge E_{Aj} & \int E_B^i \wedge E_B^j
\end{array}\) &=& 4\pi^2 \(\begin{array}{cc}
0 & \delta_i^j\\
-\delta^i_j & 0
\end{array}\)
\eea
where integration is over $M$. This can be seen as follows,
\bea
\int_M E_{Ai} \wedge E_B^j = \sum_i \int_{a^i} E_{Ak} \int_{b_i} E_B^l = 4\pi^2 \sum_i \delta^i_k \delta^l_i = 4\pi^2 \delta^l_k
\eea

Given such a symplectic basis, we define the period matrix $Z^{ij} = X^{ij} + i Y^{ij}$ by \cite{Henningson:1999dm}
\ben
E_B^i &=& X^{ij} E_{Aj} + Y^{ij} * E_{Aj}\label{period}
\een
After a holomorphic factorization, the classical contribution to the partition function is found to be
\ben
Z_{classical,6d}(Z) &=& \Theta(Z)\label{6d}
\een
where 
\bea
\Theta(Z) &=& \sum_{n_i \in {\mb{Z}}} \exp\{i\pi n_i Z^{ij} n_j\}
\eea
is the Jacobi theta function with zero characteristics. Holomorphic factorization alone does not pick the zero characteristics for us though, but this is the one that will be the focus of our interest in this Letter.

We now apply this formalism to the case when the six-manifold is a six-torus with coordinates $x^{\M} \in [0,2\pi]$. If we select one coordinate that we denote $x^{\underline{6}} := \psi$, then we have a natural metric independent choice for the symplectic basis
\bea
E_{A,MN} &=& \frac{1}{6.4\pi^2}\epsilon_{MNPQR} dx^P \wedge dx^Q \wedge dx^R\cr
E_B^{MN} &=& \frac{1}{4\pi^2}dx^M \wedge dx^N \wedge d\psi
\eea
where $\epsilon_{MNPQR} = \epsilon_{MNPQR\psi}$ is totally antisymmetric, equal to $\pm 1$. The indices runs as $M=1,2,3,4,5$ and $\M=1,2,3,4,5,\psi$. Not all these components are independent since $E_{A,MN} = - E_{A,NM}$. We may restrict ourselves to $M < N$, and then we find $b_3(T^6) = 10$ independent components in $E_{A,MN}$, and likewise for $E_B^{MN}$. The metric on a flat six-torus can be expressed in the Kaluza-Klein form as
\bea
g_{\M\N} &=& \(\begin{array}{cc}
G_{MN} + R^2 V_M V_N & R^2 V_M\\
R^2 V_N & R^2
\end{array}\)
\eea
and the inverse is
\bea
g^{\M\N} &=& \(\begin{array}{cc}
G^{MN} & -V^M\\
-V^N & \frac{1}{R^2} + V^2
\end{array}\)
\eea
where $V^2 := G^{MN} V_M V_N$. We will refer to $V_M$ as the graviphoton field. The six- and five-dimensional square-root determinants are related as
\bea
\sqrt{g} &=& \sqrt{G} R
\eea
The period matrix can be computed from the definition (\ref{period}), which here explicitly reads
\bea
E_B^{MN} &=& \frac{1}{2}\(X^{MN,PQ} E_{A,PQ} + Y^{MN,PQ} * E_{A,PQ}\)
\eea
where the important factor of $\frac{1}{2}$ arises by summing over all $P,Q$ rather than restricting the sum to $P<Q$. Then we also have
\bea
\sum_{M<N, P<Q} n_{MN} Z^{MN,PQ} n_{Pq} &=& \frac{1}{4}\sum n_{MN} Z^{MN,PQ} n_{PQ}
\eea
Following the steps in the Appendix of \cite{Gustavsson:2000kr}, we compute this period matrix with the result 
\ben
Z^{MN,PQ} &=& - G \epsilon^{MNPQR} V_R + \frac{2i\sqrt{G}}{R} G^{MN,RS}\label{periodT6}
\een

We now turn to Maxwell theory on euclidean $T^5$ with metric $G_{MN}$ and Action 
\ben
S[F] &=& \frac{1}{8\pi^2 R} \int F \wedge * F + \frac{i}{8\pi^2} \int V \wedge F\wedge F\label{proposed}
\een
One argument for having this normalization of the Maxwell term can be found in \cite{Lambert:2010iw}. The second term, which we may refer to as the graviphoton term as it involves $V = V_M dx^M$, was also discussed in \cite{Witten:1998uka} and we will shortly give two arguments why it should be normalized in precisely this way. Hodge decomposition in combination with the Bianchi identity gives us 
\bea
F &=& F^0 + dA
\eea
where $A$ is globally defined and $F^0$ is a harmonic two-form. On a five-torus we have the harmonics
\bea
F^0_{MN} &=& \frac{1}{2\pi} n_{MN}
\eea
where the $n_{MN}$ are integer quantized by the Dirac charge quantization. It is now easy to see that we can indeed identify the classical part of the six-dimensional partition function with the classical part of the five-dimensional partition function,
\bea
Z_{classical,5d} &=& Z_{classical,6d}
\eea
where
\bea
Z_{classical,5d} &=& \sum_{n_{MN}}\exp \{-S[F^0]\}
\eea
and $Z_{classical,6d}$ is given by (\ref{6d}), (\ref{periodT6}).

For the supersymmetric M5 brane the quantum fluctuations cancel between bosons and fermions. With periodic boundary conditions, the partition function in the Hamilton formalism is given by $Z = \tr((-1)^F e^{i 2\pi H})$ where $H$ is the Hamiltonian. All the quantum fluctuations have positive energy and there is a finite size energy gap to the lowest excited state due to the finite size of the six torus which renders all momenta to be integer quantized. The same is true for the supersymmetric D4 brane. We conclude that the partition functions for M5 and D4 are exactly the same in the Abelian case when the six-manifold is a flat six-torus. The corresponding five-manifold may be taken as any of the five-tori that one can embed into the six-torus with the graviphoton field and five-dimensional metric defined accordingly.

We end this Letter by providing another argument for the presence of the graviphoton term in the five-dimensional Action, which also gives a justification of the normalization constant that we have chosen. Let us consider a $4+2$ split of the six-manifold and restrict the metric to be on the form
\bea
g_{\M\N} &=& \(\begin{array}{ccc}
g_{\mu\nu} & 0 & 0\\
0 & g_{55} & R^2 V_5\\
0 & R^2 V_5 & R^2
\end{array}\)
\eea
Let us define
\bea
\(\begin{array}{cc}
g_{55} & R^2 V_5\\
R^2 V_5 & R^2 
\end{array}\) &=& 
R^2 \(\begin{array}{cc}
|\tau|^2 & \tau_1\\
\tau_1 & 1
\end{array}\)
\eea
with\footnote{We use the convention $\frac{1}{4g^2} F_{MN} F^{MN}$ to define $g$. This means that we relate our $g^2$ to $\frac{1}{2}g^2$ in \cite{Witten:1995gf} which uses the convention $\frac{1}{2g^2} F_{MN} F^{MN}$ instead.}
\bea
\tau = \tau_1 + i \tau_2 = \frac{\theta}{2\pi} + \frac{2\pi i}{g^2_{4d}}
\eea
The graviphoton is then given explicitly by
\bea
V_M &=& \frac{\theta}{2\pi} \delta_M^5
\eea
We now dimensionally reduce the five-dimensional Action (\ref{proposed}) along $x^5$ to four dimensions where we get
\bea
\frac{1}{4g^2_{4d}} \int d^4 x \sqrt{g} F_{\mu\nu} F^{\mu\nu} - \frac{i\theta}{32 \pi^2} \int d^4 x g \epsilon^{\mu\nu\lambda\tau} F_{\mu\nu} F_{\lambda\tau}
\eea
Here
\bea
g^2_{4d} &=& \frac{2\pi}{|\tau|^2}
\eea
is the dimensionless four-dimensional coupling constant. This four-dimensional Action has been derived from M5 brane by dimensional reduction \cite{Verlinde:1995mz}. 

Let us finally comment on a $3+3$ split and consider a six-manifold on the form ${\mb{R}}^{1,2} \times S^3/{\mb{Z}}_k$. Here ${\mb{Z}_k}$ acts on $z^a \in {\mb{C}}^2$ as $z^a \rightarrow e^{\frac{2\pi i}{k}} z^a$, and $S^3$ consists of the points which satisfy $z^1 \bar{z}^1 + z^2 \bar{z}^2 = R^2$. In this case the graviphoton will be a connection one-form on $S^2$ base-manifold of $S^3/{\mb{Z}_k}$ which obeys
\bea
\int_{S^2} W &=& 2\pi k
\eea
where, locally, $W = dV$, and the ${\mb{Z}}_k$ acts on the $S^1$ fiber only. To see this more clearly, let us write down the metric on $S^3$. If we denote the fiber coordinate by $\psi$ then orbifolding amounts to the identification $\psi \sim \psi + \frac{2\pi}{k}$. This leads us to rewrite the Kaluza-Klein metric in the form
\bea
ds^2 &=& G_{mn} d\sigma^m d\sigma^n + \(\frac{R}{k}\)^2 \(k d\psi + k V\)^2
\eea
where $G_{mn}$ is the metric on $S^2$ and where we scale the radius of the fiber by $\frac{1}{k}$ at the same time as we scale the graviphoton by $k$. Then we rename $\t \psi = k \psi$, which after orbifolding is $2\pi$ periodic. Then we see that also $\t V = k V$ is subject to the above Dirac quantization (which should now be written in terms of tilde variables, but which we now remove for notational convenience). Let us now consider dimensional reduction on $S^2$. In this situation we get, upon an integration by parts,
\bea
\frac{1}{32 \pi^2} \int_{{\mb{R}}^{1,2} \times S^2} d^5 x G\epsilon^{MNPQR} V_R F_{MN} F_{PQ} &=& \frac{k}{4\pi} \int_{{\mb{R}^{1,2}}} d^3 x \epsilon^{\mu\nu\lambda} A_{\mu} \partial_{\nu} A_{\lambda}
\eea
where we split $M = (\mu,m)$. This now provides a clean connection between the $k$ of the orbifold $S^3/{\mb{Z}_k}$ and the $k$ of the three-dimensional Chern-Simons term. Consistency of the Chern-Simons term for arbitrary $k\in \mb{Z}$ also shows that the normalization of the five-dimensional graviphoton term has to be the way we have chosen it.

\subsubsection*{Acknowledgements}
I would like to thank Soo-Jong Rey for discussions. This work was supported by NRF Mid-career Researcher Program 2011-0013228.

\end{document}